\documentclass[journal]{IEEEtran}
\usepackage{cite,graphicx, psfrag, subfigure, url, amsmath, amssymb, array,color }

\def \Diag  {\operatorname{Diag}}
\def \tr    {\textrm{tr}}
\def \SNR   {\operatorname{SNR}}
\def \BER   {\operatorname{BER}}

\def \MSE   {\mathbf{E}}

\def \I     {{\mathbf I}}

\def \H     {{\mathbf H}}
\def \E     {{\mathbf E}}

\def \G     {{\mathbf G}}
\def \P     {{\mathbf P}}
\def \A     {{\mathbf A}}
\def \B     {{\mathbf B}}
\def \C     {{\mathbf C}}

\def \V     {{\mathbf V}}
\def \U     {{\mathbf U}}

\def \N     {{\mathbf N}}

\def \Z     {{\mathbf Z}}
\def \R     {{\mathbf R}}

\def \L     {{\mathbf L}}
\def \R     {{\mathbf R}}
\def \Q     {{\mathbf Q}}

\def \Lambdab   {{{\mathbf \Lambda}}}

\def \Ptot  {P_{\textrm{total}}}
\def \Lff   {\mathbf{L}_{1 1}}
\def \Lkk   {\mathbf{L}_{K K}}
\def \Lii   {\mathbf{L}_{i i}}

\def \Lall  {{\mathbf{L}_{1 1}}, \dots,{\mathbf{L}_{K K}}}

\def \sign  {\sigma_{n}^{2}}

\def \0     {{\mathbf {0}}}
\def \e     {{\mathbf e}}
\def \s     {{\mathbf s}}

\def \e     {{\mathbf e}}
\def \x     {{\mathbf x}}
\def \y     {{\mathbf y}}
\def \n     {{\mathbf n}}
\def \b     {{\mathbf b}}

\def \e     {{\mathbf e}}

\def \a     {{\mathbf a}}
\def \b     {{\mathbf b}}

\def \l     {{\boldsymbol{\mit{l}}}}
\def \lb    {\overline{ \l }}

\def \sh    {\hat{\mathbf{s}}}

\def \bP    {\mathbf{\overline{P}}}

\def \Q     {\mathbf {Q}}

\newtheorem{lem}{Lemma}
\newtheorem{theorem}{Theorem}

\newtheorem{definition}{Definition}

\newcommand{\rev}[1]{{#1}}
\begin{document}
\title{A Design Framework for Limited Feedback MIMO Systems with Zero-Forcing DFE}
\author{Michael Botros Shenouda and
        Timothy N. Davidson,~\IEEEmembership{Member,~IEEE,}
\thanks{
Manuscript received 4 November 2007; revised 15 April 2008.
This work was supported in part by the Natural Sciences and Engineering Research Council of Canada and
an Ontario Graduate Scholarship in Science and Technology.
The work of the second author is also supported by the Canada Research Chairs Program.
A preliminary version of this manuscript appears in \textit{Proc. Canadian Wkshp Inform. Theory},  Edmonton, June 2007.
} 
%
\thanks{
The authors are with the Department of Electrical and Computer Engineering,
McMaster University, Hamilton, Ontario, Canada.
(botrosmw,davidson)@mcmaster.ca
}
}
%
\maketitle

\begin{abstract}
We consider the design of multiple-input multiple-output communication systems with a linear precoder at the transmitter,  zero-forcing decision feedback equalization (ZF-DFE) at the receiver, and a low-rate feedback channel that enables communication from the receiver to the transmitter.
The channel state information (CSI) available at the receiver is assumed to be perfect, and based on this information the receiver selects a suitable precoder from a codebook and feeds back the index of this precoder to the transmitter.
Our approach to the design of the components of this limited feedback scheme is based on the development, herein, of a unified framework for the joint design of the precoder and the ZF-DFE   under the assumption that perfect CSI is available at both the transmitter and the receiver.
The framework is general and embraces a wide range of design criteria.
This framework enables us to characterize the statistical distribution of the optimal precoder in a standard Rayleigh fading environment.
Using this distribution, we show that codebooks constructed from Grassmann packings minimize an upper bound on an average distortion measure, and hence are natural candidates for the codebook in limited feedback systems.
%
%
Our simulation studies show that the proposed limited feedback scheme can provide significantly better performance at a lower feedback rate than existing schemes in which the detection order is fed back to the transmitter.
\end{abstract}


\section{Introduction}
Multiple-input multiple-output (MIMO) communication schemes offer the potential for significant increases in spectral efficiency over their single-input single-output counterparts by enabling  simultaneous transmission of independent data streams.
MIMO schemes also offer the potential for significant performance gains in a variety of other metrics.
Standard transceiver architectures for these schemes include linear precoding and equalization, and the combination of linear precoding and decision feedback equalization (DFE), which offers the potential for improved performance over the linear approach while maintaining comparable complexity.
For scenarios in which accurate channel state information (CSI) is available at both the transmitter and the receiver, there is a well established framework that unifies the design of linear transceivers under many design criteria \cite{Palomar_2003}.
A counterpart for the design of systems with DFE has recently emerged \cite{Botros_2007_ICASSP,Botros_THP_DFE_JSAC,Jiang_2007_DFE,Palomar_Book}.
This framework was also extended to MIMO systems with pre-interference subtraction at the transmitter in \cite{Botros_2007_ICASSP}.
%
However, in many scenarios, such as frequency division duplex systems, obtaining accurate CSI at the transmitter may require a considerable amount of feedback to the transmitter.
An approach that allows the designer to limit the required amount of the feedback  is to quantize the transmitter design.
In these limited feedback schemes \cite{Love_Value-FB_2004}, the receiver uses its CSI to choose the best transmitter design from a codebook of available designs,  and then feeds back the index of this precoder  to the transmitter.
This strategy has been considered for beamforming schemes (e.g., \cite{Narula_1998, Visotsky_FR-beamforming,Love_2003_Grass-BF,Mukkavilli_2003_FR-Beamforming, Santipach_2003_FR-Beamforming,Xia_2005_FR-BF2,Roh_2006_FR-Beamforming}), unitary precoding with linear equalization (e.g., \cite{Love_2005_Grass-LP}).
and unitary precoding for orthogonal space time block codes \cite{Love_2005_Grass-STBC,Jongren_2004_FR-STBC}.
For zero-forcing DFE schemes,  a limited feedback scheme in which the receiver feeds back  the order of interference cancellation was  proposed in \cite{Bae_2006_FR-ZFDFE,Jiang_2006_FR-ZFDFE}.

In this work, we consider the design of a limited feedback scheme for systems with a (general) linear precoder at the transmitter and zero-forcing DFE at the receiver.
Our designs are based on a unified framework, developed herein, for the joint design of the precoder and the ZF-DFE in the presence of perfect CSI.
This framework embraces a wide range of design criteria that can be expressed as functions of the mean square error (MSE) of each data stream, including minimization of the total MSE, minimization of the average bit error rate (BER), and maximization of the Gaussian mutual information.
In particular, we show that the optimal precoder for systems with a zero-forcing DFE is the same for all these criteria; a property that cannot be achieved by a linear transceiver.
Furthermore, we show that the optimal precoder for these objectives is a scaled unitary matrix that is isotropically distributed (over the  Stiefel manifold of unitary matrices).
Using this distribution, we show that codebooks constructed from Grassmann subspace packings minimize an upper bound on an average distortion measure, and hence are excellent candidates for the codebook in limited feedback schemes for systems with zero-forcing DFE.
In contrast, the application of Grassmann codebooks in limited feedback schemes with linear receivers (e.g., \cite{Love_2005_Grass-LP}) involves an inherent compromise, because the optimal precoder in the presence of perfect CSI and a total power constraint is not unitary.
Since the scheme that we propose involves the construction of codebooks for isotropically distributed unitary matrices, our scheme subsumes that in \cite{Bae_2006_FR-ZFDFE,Jiang_2006_FR-ZFDFE}, in which the precoder is, by construction, a permutation matrix.
Our simulation studies suggest that the additional degrees of freedom available in our approach enable  our scheme to provide significantly better performance than that in \cite{Bae_2006_FR-ZFDFE,Jiang_2006_FR-ZFDFE} while using a lower feedback rate.

Our notation is as follows:
Boldface type is used to denote matrices and vectors;
$\a_i$ denotes the $i^{\text{th}}$ element of the vector $\a$,
$\A_{i j}$ denotes the element at the intersection  of the $i^{\text{th}}$ row and $j^{\text{th}}$ column of the matrix $\A$,
$\A^H$ denotes the conjugate transpose of $\A$, and $(\A)^{\dagger}$ denotes the (minimum norm) pseudo-inverse of $\A$.
The terms $\det(\A)$, and $\| \A \|_{2}$ denote  the  determinant and the two-norm (maximum singular value) of $\A$, respectively.
The notation $\operatorname{Diag}(\x)$ denotes the diagonal matrix whose elements are the elements of~$\x$.

\section{System Model}
We consider a point-to-point communication system with $N_t$ transmit antennas and $N_r$ receive antennas
that transmits $K$ data streams simultaneously, where $K$ is no greater than the rank of the channel matrix $\H$.
\rev{
We adopt a narrow band block fading channel model, and we consider MIMO communications systems that use (generalized) zero-forcing decision feedback equalization, e.g., \cite{Cioffi_1979_GDFE,Xu_2006_DFE}, for spatial equalization.}
At the transmitter, the input data vector $\s \in \mathbb{C}^{K}$ is linearly precoded using $\P$ to generate the transmitted data vector $\x \in \mathbb{C}^{N_t}$,
\begin{equation}
\x = \P \s.
\label{Grass_s}
\end{equation}
Without loss of generality, we will assume that $\mathrm{E}  \{\s \s^H\} = \I$, and hence the
total transmitter power constraint can be written as $\mathrm{E} \{\x^H \x\} = \tr (\P^H \P) \le
\Ptot$.

\begin{figure}
\centering
\includegraphics[width=3.5in]{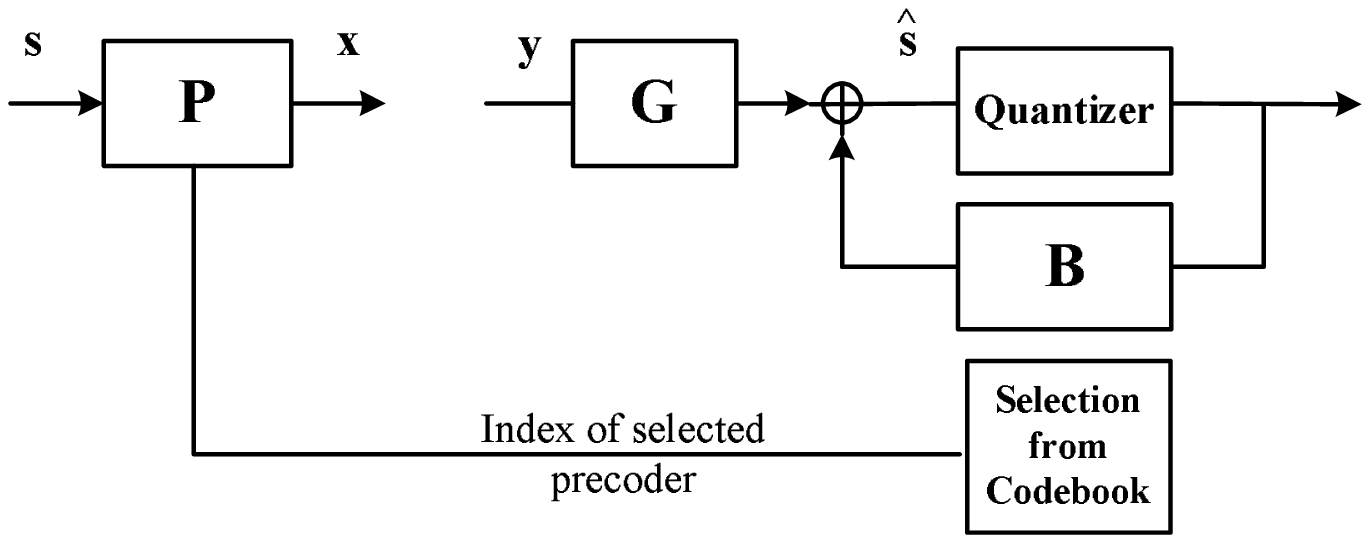}
\caption{MIMO transceiver with DFE using limited feedback.}
\label{DFE}
\end{figure}

The vector of received signals is given by
\begin{equation}
\y = \H \P \s + \n,
\end{equation}
where $\H$ is the channel matrix and $\n$ is the vector of  additive noise which is assumed to have zero-mean and a covariance
matrix $\mathrm{E}\{\n \n^H\} = \sign \I$.
As illustrated in Fig.~\ref{DFE}, following linear processing  using the feedforward matrix $\G$, the receiver
makes successive decisions on each symbol  by subtracting the effect of previously decided symbols.
Hence, the feedback  matrix $\B$ is strictly lower triangular.
This system model embraces linear precoding and equalization as a special case when $\B = \0 $.
Assuming correct previous decisions, the vector of inputs to the quantizer is given by
\begin{equation}
\sh  = (\G \H \P - \B)\s + \G \n.    \label{Grass_DFE_channel}
\end{equation}
By defining the error signal $\e = \s - \sh$, the  error covariance matrix (the ``MSE" matrix) can be written
as
\begin{multline}
\textstyle
\MSE =
\mathrm{E}\{\e \e^H\} =
\C\C^H  - \C \P^H \H^H \G^H     - \G   \H  \P \C^H\\
+ \G  \H \P \P^H \H^H \G^H      +  \sign \G    \G^H,   \label{Grass_DFE_MSE}
\end{multline}
where $\C = \I + \B$ is a unit diagonal lower triangular matrix.

We will consider communication schemes in which perfect CSI is available only at the
receiver.
Based on its channel knowledge, the receiver selects a suitable precoding
matrix from a codebook of precoders  $\mathcal{P}$ of size $|\mathcal{P}|$, and feeds that index
back to the transmitter using $\log_2 |\mathcal{P}|$ information bits; see Fig~\ref{DFE}.
In order to develop effective methods for quantizing the precoding matrix, we first need  to
characterize the optimal precoding matrix for different design criteria in the presence of perfect CSI.
We will then use the statistical distribution of this optimal precoder to define the distortion measures that are required to design the codebook for the limited feedback scheme.

\section{Unified Framework for Zero-Forcing DFE} \label{Grass_Sec_ZF-Framework}
In this section, we develop a general framework for the joint design of the transceiver matrices $\G, \C = \I + \B$, and $\P$ in the presence of perfect CSI.
We consider system design criteria that are expressed as functions of the (logarithm of the)
MSE of the individual data streams $\E_{ii}$.
The proposed framework embraces a wide range of design objectives.
\rev{It includes objectives for which  optimal designs are already available (e.g., the total MSE, \cite{Xu_2006_DFE}), and several other objectives for which the optimal transceiver design has remained an open problem.}
The framework can be regarded as a counterpart for the existing framework of linear transceiver  design \cite{Palomar_2003}.
Here, the framework is derived for DFEs with a zero-forcing constraint, but  an analogous framework can be developed in the absence of this constraint \cite{Botros_2007_ICASSP,Botros_THP_DFE_JSAC,Jiang_2007_DFE,Palomar_Book}.
\subsection{ZF-DFE Receiver Design}
The zero-forcing design criterion implies
\begin{equation}
\G \H \P - \B = \I.
\label{DFE_ZF_criteria}
\end{equation}
Given the assumption that $K \le \text{rank}(\H)$, the condition in (\ref{DFE_ZF_criteria}) can be achieved so long as $\P$ is chosen such that $\text{rank}(\H\P) = K$. In that case, the feedforward matrix $\G$ is given by
\begin{equation}
\G = \C (\H \P)^{\dagger}.
\label{DFE_ZF_G}
\end{equation}
Since $\H \P$ has full column rank, the pseudo-inverse in (\ref{DFE_ZF_G}) can be written as
\begin{equation}
(\H \P)^{\dagger}  = (\P^H \H^H \H \P)^{-1} \P^H \H^H.
\end{equation}
Using the expression for $\G$ in (\ref{DFE_ZF_G}), the MSE matrix in (\ref{Grass_DFE_MSE}) reduces to
\begin{equation}
\E =  \C \N \C^H,
\end{equation}
where $\N = \sign (\P^H \H^H \H \P)^{-1}$ is a positive definite Hermitian matrix.
The optimal matrix $\C$, that minimizes the MSE of each individual data stream, subject to being unit diagonal and lower triangular, is given by \cite{Botros_2007_ICASSP}
\begin{equation}
\C = \Diag \left( \Lall \right) \L^{-1},
\label{DFE_ZF_C}
\end{equation}
where $\N = \L \L^H$ is the Cholesky factorization of $\N$, and $\L$ is a lower triangular matrix with strictly positive diagonal entries.
Using this optimal $\C$, the MSE matrix can be  rewritten as
\begin{equation}
\MSE  = \Diag  \left(  \Lff^2, \ldots, \Lkk^2 \right),
\end{equation}
where $\Lii$ is the $i^{\text{th}}$ diagonal element of $\L$. Hence, the SNR of each data stream is
\begin{equation}
\SNR_k = \frac{1}{\MSE_{kk}} = \frac{1}{\L_{kk}^2}.
\end{equation}
\subsection{Transmitter Design}
Given the optimal   $\G$ and $\C$, our next step is to design a precoding matrix $\P$ so as to
optimize design criteria that are expressed as functions of the (logarithm of the)  MSE of each individual stream.
To derive the optimal precoding matrix, we will first obtain some inequalities that involve the logarithm of the MSE of the individual data streams,
\begin{equation}
\l =(\ln \Lff^2, \ldots, \ln \Lkk^2),
\label{Gras_def_l}
\end{equation}
using concepts from majorization theory.

\begin{definition}[Additive Majorization \cite{Marshall_1979}]
Let $\a, \b \in \mathbb{R}^{K}$ and let $a_{[1]}, \ldots, a_{[K]}$ denote the re-ordering of the elements of $\a$ in a
non-\rev{increasing} order; i.e., $a_{[1]} \ge \ldots \ge a_{[K]}$.
The vector $\b$ is said to majorize $\a$,
$\a~\prec~\b$, if
\begin{align}
\sum_{i=1}^{j}   \a_{[i]}       &   \le    \sum_{i=1}^{j}    \b_{[i]} \qquad \text{for} \; j = 1, \ldots, K-1,  \\
\sum_{i=1}^{K}   \a_{[i]}       &    =     \sum_{i=1}^{K} \b_{[i]}.
\end{align} \hfill $\Box$
\end{definition}
The following lemma will play a key role in our framework.
\begin{lem}
\label{Grass_Lemma_majorization_ineq}
For the Cholesky factorization $\N = \L \L^H$, the following inequalities hold:
\begin{equation*}
 \frac{\ln\det(\N)}{K} (1, \ldots, 1) \; \prec \; \l \; \prec \; (\ln
\lambda_1(\N),~\ldots,~\ln \lambda_K(\N)),
\end{equation*}
where $\lambda_k(\N)$ is the $k^{\text{th}}$ largest eigen value of $\N$.
\hfill   $\Box$
\end{lem}

\begin{proof}
To prove the first inequality, we observe that any vector $\a~\in~\mathbb{R}^{K}$
majorizes its mean vector $\overline{\a}$, whose elements are all equal to the mean $\overline{\a}_i~=~\frac{1}{K}~ \sum_{i = 1}^{K}~\a_i$. That is,
\begin{equation}
\overline{\a}  \prec \a.
\end{equation}
Since $\N = \L \L^H$, we have that $ \prod_{k=1}^{K} \L_{kk}^2 = \det (\L \L^H  ) = \det
(\N)$. Hence, the first inequality follows directly.
The second inequality follows by applying Weyl's inequality \cite{Weyl_1949} to
the matrix $\L$.
\end{proof}
It is worth observing that the second inequality in Lemma~\ref{Grass_Lemma_majorization_ineq} holds with equality when $\L$ is normal
\cite{Weyl_1949}.
Since $\L$ is a lower triangular matrix,  in order to be normal it must be a diagonal
matrix \cite{MAtrix_Analysis}. If $\L$ is diagonal, the matrix $\C$ will then be equal to $\I$ and decision feedback
equalization will reduce to linear equalization.

The proposed designs will be based on the following classes of functions \cite{Marshall_1979}.
\begin{definition}[Schur-convex and Schur-concave functions]
A real-valued function $f(\x)$ defined on a subset $\mathcal{A}$ of $\mathbb{R}^{K}$ is said to be Schur-convex if
\begin{equation}
\a \prec \b   \:   \textrm{on} \:  \mathcal{A}  \Rightarrow f(\a) \le f(\b),
\end{equation}
and is said to be Schur-concave if
\begin{equation}
 \a \prec \b   \:   \textrm{on} \:  \mathcal{A}  \Rightarrow f(\a) \ge f(\b).
\end{equation}
\hfill $\Box$
\end{definition}
In particular, we will consider communication objectives that can be expressed as the minimization of increasing  functions of the MSEs of each data stream, $g(\Lff^2, \ldots, \Lkk^2) = g (e^{\l_1}, \ldots, e^{\l_K}) = g(e^{\l})$, that are either Schur-convex or Schur-concave functions of the logarithm of the MSEs, $\l$.

Let $\H^H \H = \U \Lambdab_{\H}\U^H$ be the eigen value decomposition of  $\H^H \H$ such
that the entries of the  diagonal matrix  $\Lambdab_{\H}$  are squared singular values of $\H$,
$\sigma^2_k(\H)$, in descending order.
Let $\U_{1}$ and $\Lambdab_{\H 1}$  be the first $K$ columns of $\U$ and
$\Lambdab_{\H}$, respectively.
The optimal precoders for the above two classes of design criteria are given by the
following theorem.

\begin{theorem}
\label{Grass_Theorem1}
The optimal precoder for the class of objectives for which $g(e^{\l})$   is a Schur-convex function
of the logarithm of the MSEs is independent of the actual form of $g(\cdot)$ and is given by:
\begin{equation}
\P = \sqrt{ \frac{\Ptot}{K}} \U_1 \V(\Lambdab_{\H 1}), \label{Grass_Theorem_1}
\end{equation}
where $\V(\Lambdab_{\H 1})$ is a  unitary matrix   that results  in the QR decomposition of $\Lambdab_{\H
1}^{-1/2} \V(\Lambdab_{\H 1}) = \Q \R$ having an $\R$ factor with equal diagonal elements.\\
For the class of objectives for which $g(e^{\l})$ is a Schur-concave function
of the logarithm of the MSEs, the optimal solution results in $\B = \mathbf{0} $, and hence the optimal zero-forcing linear transceiver is an optimal transceiver for a system with a zero-forcing DFE.
\end{theorem}
\begin{proof}
See the Appendix.
\end{proof}
\rev{ Algorithms for obtaining a matrix $\mathbf{\Phi}$ such that the R-factor of the  QR  decomposition of $\A \mathbf{\Phi}$  has equal diagonal elements were introduced in  \cite{Zhang_2005_QRS,Jiang_2005_GMD_math}, and $\V$  in (\ref{Grass_Theorem_1}) can be obtained by applying the algorithms therein  to the matrix  $\Lambdab^{-1/2}_{\H 1}$.}

As illustrated by the following examples, the developed framework embraces a wide range of design criteria:
\begin{itemize}
\item
\textit{Minimization of the sum of the individual MSEs:}
In this case the objective is to minimize
\begin{equation}
g(e^{\l}) =  \sum_{k = 1}^{K}  e^{\l_k}.
\end{equation}
Here, $g(e^{\l})$  takes the form $ \sum_{k= 1}^{K} h(\l_k)$ for the convex function $h(\l_k) = e^{\l_k}$, and
hence it is a Schur-convex function of $\l$, \cite{Marshall_1979}.
\item
\textit{Minimization of the maximum MSE / Maximization of minimum SNR:}
In this case the objective is to minimize
\begin{equation}
g(e^{\l}) = \max_k (e^{\l_k}),
\end{equation}
which is the composition of the increasing Schur-convex function $\max_k \x_k$ with the increasing and
convex function  $e^{\l_k}$. Hence, $g(e^{\l})$ is also a Schur-convex function
\cite[pp. 63]{Marshall_1979}.
\item
\textit{Minimization of the average Bit Error Rate:}
This corresponds to the minimization of the objective
\begin{equation}
g(e^{\l}) = \frac{1}{K}  \sum_{k = 1}^{K}  \BER(\SNR_k) =  \sum_{k = 1}^{K}  \BER(e^{-\l_k}),
\end{equation}
where the $\BER$ expression will depend on the constellation used, and we have assumed that the same constellation is used for each element of $\s$ in (\ref{Grass_s}).
It can be verified that under a mild constraint on the SNR, the BER expressions for BPSK and M-QAM constellations are convex functions of $\l_k$. Hence, $g(e^{\l})$ is a Schur-convex function of $\l$.
\item
\textit{Maximization of Gaussian mutual information}
This corresponds to the minimization of
\begin{equation}
g(e^{\l}) =  \sum_{k = 1}^{K} - \log( 1 + e^{-\l_k}),
\label{Grass_GuassianMutualInfo_eg}
\end{equation}
which takes the form $ \sum_{k= 1}^{K} h(\l_k)$ for the convex function $h(\l_k) = - \log( 1 + e^{-\l_k}) $, and
hence it is a Schur-convex function of $\l$.
\item
\textit{Minimization of the product of MSEs:}
Minimization of the product of the individual MSEs (or equivalently, the geometric mean of the MSEs) corresponds to the minimization of
\begin{equation}
g(e^{\l}) = \log \prod_{k = 1}^{K} e^{\l_k} =  \sum_{k = 1}^{K} \l_k,
\end{equation}
which is both Schur-convex and Schur-concave.
Furthermore, since  $\sum_{k = 1}^{K} \l_k = -\sum_{k = 1}^{K} \log(\SNR) $,
at high SNR the minimization of the product of the MSEs corresponds to the maximization of the Gaussian mutual information.
\end{itemize}

As demonstrated by Theorem~\ref{Grass_Theorem1} and the above examples, the optimal precoder for a system with zero-forcing DFE and a design objective from the Schur-convex class simultaneously optimizes the total MSE, the average bit error rate, and the Gaussian mutual information.
MIMO systems with linear precoding and equalization do not achieve this simultaneous optimality, and in the general case each of these objectives results in a  different optimal precoder  \cite{Palomar_2003}.
For design criteria that can be expressed as the minimization of objectives that are both Schur-convex and Schur-concave, both the optimal Schur-convex design in (\ref{Grass_Theorem_1}) and the optimal linear transceiver will yield the same objective value.
In the following sections, we will consider the efficient design of codebooks for limited feedback systems with Schur-convex objectives. Our first step will be to obtain the statistical distribution of the optimal precoder.

\section{Statistical Distribution of Optimal Precoder for Schur-Convex Objectives}
The optimal precoder for the Schur-convex class of objectives can be written as
\begin{equation}
\P = \sqrt{\frac{\Ptot}{K} } \: \bP,
\label{Grass_opt_P}
\end{equation}
where the matrix $\bP = \U_1 \V (\Lambdab_{\H 1}) $ belongs to the Stiefel manifold
$\mathcal S(N_t, K)$ of complex $N_t \times K$ matrices with orthonormal columns.
The statistical distribution of $\bP$ in (\ref{Grass_opt_P}) plays a key role in the design of the codebooks, and is established in Theorem~2 below. First, we establish an intermediate result.
\begin{lem}
For an i.i.d.\ Rayleigh fading channel matrix $\H$, the matrices $\U_1$ and $\V(\Lambdab_{\H 1})$ are
statistically independent. Furthermore, $\U_1$ is isotropically distributed over the manifold $\mathcal
S(N_t, K)$.
\label{Grass_Lemma2}
\end{lem}
\begin{proof}
The proof follows directly from the isotropic distribution of the eigen vectors of the Wishart distributed matrix $\H^H \H$ and its independence of the eigen values.
\end{proof}

\begin{theorem}
For an i.i.d.\ Rayleigh fading channel matrix $\H$, the normalized optimal precoder
matrix $\bP$ is isotropically distributed over the Stiefel manifold  $\mathcal S(N_t,
K)$.
\end{theorem}
\begin{proof}
We first observe from Lemma~\ref{Grass_Lemma2} that $\U_1$ is isotropically distributed over the manifold $\mathcal
S(N_t, K)$. Hence, its probability distribution $p(\U_1)$ is unaffected by
post-multiplication by any \textit{deterministic} unitary matrix $\Z$; i.e., $p(\U_1) =
p (\U_1 \Z)$. Hence,
\begin{eqnarray}
p(\bP) & = &  \int p( \bP  | \V ) \: p(\V) \: d\V \\
       & = &  \int p( \U_1      ) \: p(\V) \: d\V  = p (\U_1),
\end{eqnarray}
Since $\U_1$ is isotropically distributed, then so is $\bP$.
\end{proof}

It is worth noting that for MIMO systems with linear precoding and equalization, the optimal precoder will not be isotropically distributed.
That is true for a wide range of objectives under a total power constraint (e.g., \cite{Palomar_2003} and the references therein), and holds for both zero-forcing and MMSE linear receivers.
That said, some quantization methods for linear transceivers have been based on a suboptimal underlying scheme that selects the best unitary precoding matrix; e.g., \cite{Love_2005_Grass-LP}.
In that case the distribution of the unquantized precoder is isotropic.
In the case of systems with a zero-forcing DFE, we have shown that selection of the best unitary precoding matrix is optimal.

\section{Precoder Selection and Codebook design}
In order to study the codebook design problem, we will first consider the selection method
for choosing the best precoding matrix from a given codebook $\mathcal{P}$.
\subsection{Precoding Matrix Selection}
Given a codebook for quantizing the normalized optimal precoding matrix
$\bP$, $\mathcal{P} = \{ \bP^j, j = 1, \ldots , |\mathcal{P}| \}$, and a cost function $g(\cdot)$ associated with the design criterion, the receiver will select a normalized precoding matrix from the codebook that yields the minimum value for the cost function; i.e., the receiver will select the index
\begin{equation}
\arg \min_{j = 1, \ldots, |\mathcal{P}|} g(e^{\l^{j}}),
\end{equation}
where $\l^{j}$ is the vector containing the logarithm of the diagonal elements of $\L^j$, the Cholesky factor of $\N^j = \sign \big(\frac{K}{\Ptot} \bP^{j\;H} \H^H \H \bP^j\big)^{-1}$.
The quality of a given codebook can be measured in terms of the average degradation in the value of the objective that is incurred by using a precoder from the codebook rather than the optimal precoder in Theorem~1.
Borrowing terminology from the source coding literature, we will refer to this degradation, and various bounds thereon, as distortion measures for the quantization scheme.
\subsection{Grassmann Packing and Codebook Design}
In the following section we will consider the design of codebooks  to minimize distortion measures for the broad class of  objectives $g(e^\l)$ that are Schur-convex in $\l$.
As shown in the previous section, for these objectives the optimal normalized precoder is uniformly distributed over the Stiefel manifold $\mathcal S(N_t, K)$. We observe that the range of the columns of any normalized precoding matrix $\bP$ represents a $K$ dimensional subspace, $R_{\bP}$, of $\mathbb{C}^{N_t}$.
Hence, the desired codebook $\mathcal{P} = \{ \bP^j, j = 1, \ldots , |\mathcal{P}| \}$ represents a set of subspaces $\mathcal{R} = \{ R_{\bP^j}, j = 1, \ldots , |\mathcal{P}| \}$, and each of these subspaces can be represented as a point in the associated quotient space, namely the Grassmann Manifold; \rev{e.g., \cite{Edelman_1998_geometry_algo,Manton_2002_Grass}}.
In the next section, we will relate the problem of designing codebooks that minimize suitable distortion measures to the Grassmann packing problem that selects a set of subspaces  such that the minimum pairwise distance between any two subspaces in the packing  is maximized.
The distances between two subspaces $R_{\bP^1}$ and $R_{\bP^2}$ can be defined in different ways \cite{Barg_2002}. For example,  the projection 2-norm is defined as
\begin{equation}
 \text{dist}_{\text{proj2}}  ( \bP^1  ,   \bP^2 ) = \Bigl\| \bP^1 \: \bP^{1 \: H}  -  \: \bP^2 \: \bP^{2 \: H} \Bigr\|_2,
\label{projection_2_dist_def}
\end{equation}
while the Fubini-Study distance is defined as
\begin{equation}
\text{dist}_{\text{FS}} ( \bP^1  ,   \bP^2 ) =  \arccos \Bigl | \det( \bP^{1 \: H}  \: \bP^2) \Bigr |.
\label{fubini_dist_def}
\end{equation}
For a given set or a packing of subspaces and a given distance measure, we will denote the minimum pairwise distance between any two subspaces in the packing by
\begin{equation}
d   = \min_{1 \le i < j \le |\mathcal{P}| }  \text{dist} ( \bP^j  ,   \bP^i ).
\label{Grass_packing_dist_d_def}
\end{equation}
In addition to the minimum distance of the packing $d $, we will also be interested in its density $D$; e.g., \cite{Barg_2002}.
In our context, the density is the probability that the range space of an isotropically distributed unitary matrix falls within a distance $d/2$ of any of the subspaces of the packing, and is function of $d$, $|\mathcal{P}|$ and the volume of the manifold; see \cite{Barg_2002}.
In the following two sections, we will show that codebooks from certain optimized Grassmann packings minimize distortion measures that are appropriate for two subclasses of the Schur-convex objectives: the strict Schur-convex objectives, and the objectives that are both Schur-convex and Schur-concave functions of $\l$.

\subsection{Codebook Designs for Strictly Schur-convex Objectives}
In this section we will present suitable distortion measures for objectives $g(e^\l)$ that are   Schur-convex functions of $\l$ and are not Schur-concave; e.g., the sum of the MSEs, the maximum MSE and the $\BER$.
From the first principles, we can obtain the following bounds on the these objectives:

\begin{itemize}
\item
\textit{Minimization of the sum of MSE:}\\
\begin{equation}
g(e^{\l}) =   \sum_{k = 1}^{K}  e^{\l_k}
            \le K \: \max_{k} e^{\l_k}
        = \frac{K}{ \min_{k} e^{-\l_k}}.
\end{equation}

\item
\textit{Minimization of the maximum MSE / Maximization of minimum SNR:}\\
\begin{equation}
g(e^{\l}) = \max_k (e^{\l_k}) = \frac{1}{ \min_{k} e^{-\l_k}}.
\end{equation}

\item
\textit{Minimization of the average Bit Error Rate:}\\
\begin{equation}
g(e^{\l}) =     \sum_{k = 1}^{K}  \BER(e^{-\l_k})
          \le  K \:  \BER(   \min_{k} e^{-\l_k} ).
\end{equation}
\end{itemize}
We observe that each of these bounds is expressed in terms of the minimum $\SNR$ over the $K$ data streams, $\SNR_{\text{min}}= \min_k e^{-\l_k}$.

Since each of these terms is bounded by the minimum $\SNR$, a natural choice for the distortion measure for a given codebook  is the average loss in the minimum $\SNR$ that one incurs by using a normalized precoder $\bP^{\text{quant}}$  chosen from the codebook $\mathcal{P}$ instead of using the optimal normalized precoder $\bP^{\text{opt}}$. That is,
\begin{align}
\mathcal{{E}}   & =   \mathrm{E}_{\H} \Bigl\{ \SNR_{\min} (\bP^{\text{opt}})
          -  \SNR_{\min} (\bP^{\text{quant}})  \Bigr\}  \nonumber\\
                & =   \frac {\mathrm{E}_{\H}\{\sqrt[K]{\det \Lambdab_{\H 1} }\}} {\sign}
          - \mathrm{E}_{\H} \Bigl \{\ \max_{1 \le j \le |\mathcal{P}|} \min_{1\le k \le K} e^{-\l^j_k} \Bigr \},
\label{Distortion_def}
\end{align}
where (\ref{Distortion_def}) follows by observing that the optimal $\bP$ results in
$\l_k = \frac{\ln \det (\N)}{K} $ for every $k$.
Consider the second term in the distortion measure in equation (\ref{Distortion_def}).
From the definition of the majorization relation $\a~\prec~\b$, we have $\a_{[1]}  \le
\b_{[1]}$. Hence, from Lemma~1 we have
\begin{equation}
\max_{1 \le k \le K} \l_k   \le   \ln \lambda_1 (\N)  = \ln \Bigl ( \frac{\sign}{\sigma_{\min}^2 (\H \P)}  \Bigr),
\end{equation}
from which it follows that
\begin{equation}
\mathrm{E}_{\H} \Bigl \{\max_{1 \le j \le |\mathcal{P}|} \min_{1 \le k \le K} e^{-\l^j_k} \Bigr \}
\ge
\mathrm{E}_{\H} \Bigl \{\max_{1 \le j \le |\mathcal{P}|} \frac{\sigma_{\min}^2 (\H\P^j)}{\sign} \Bigr \}.
\label{sigma_min_ineq}
\end{equation}
Hence, the distortion measure in (\ref{Distortion_def}) is upper bounded by
\begin{equation}
\mathcal{{E}} \le
\frac {\mathrm{E}_{\H}\{\sqrt[K]{\det \Lambdab_{\H 1} }\}} {\sign}
-
\mathrm{E}_{\H} \Bigl \{\max_{1 \le j \le |\mathcal{P}|} \frac{\sigma_{\min}^2 (\H\P^j)}{\sign}\Bigr\}.
\label{Dist_sigma_min_ineq}
\end{equation}
When codebooks are designed from a Grassmann packing using the projection 2-norm distance in (\ref{projection_2_dist_def}), the expectation on the right hand side of (\ref{sigma_min_ineq}) satisfies  \cite{Love_2005_Grass-LP},
\begin{equation}
\mathrm{E}_{\H} \Bigl \{ \max_{1 \le j \le |\mathcal{P}|}  \sigma_{\min}^2 (\H\P^j)   \Bigr \}
\ge
\mathrm{E}_{\H} \{\sigma_K^2 (\H) \}      D_{\text{proj2}} \Bigl (1 - \frac{d^2_{\text{proj2}}}{4} \Bigr),
\label{result_from_Love}
\end{equation}
\rev{where $d_{\text{proj2}}$ is the  minimum pairwise distance of the packing (cf. (\ref{Grass_packing_dist_d_def})) for the projection 2-norm distance, and $D_{\text{proj2}}$ is the corresponding packing density; cf. \cite{Barg_2002}.}
In addition, for a given $ | \mathcal{P}| $ the right hand side of (\ref{result_from_Love}) is an increasing function of the packing  distance $d_{\text{proj2}}$.
Using the inequality in (\ref{Dist_sigma_min_ineq}), we obtain the following upper bound on the distortion:
\begin{equation}
\mathcal{{E}} \le
\frac {\mathrm{E}_{\H}\{\sqrt[K]{\det \Lambdab_{\H 1} }\}} {\sign}
-
\frac{\mathrm{E}_{\H} \{\sigma_K^2 (\H) \}}{\sign}   D_{\text{proj2}} \Bigl(1 - \frac{d^2_{\text{proj2}}}{4} \Bigr),
\label{Grass_Dist_bound-denisty}
\end{equation}
which, for a given $|\mathcal{P}|$, is a decreasing function of the packing distance $d_{\text{proj2}}$.
The bound on the right hand side of (\ref{Grass_Dist_bound-denisty}) can be easily manipulated by choosing the codebook from a Grassmann packing that is designed to maximize the packing distance $d$ in (\ref{Grass_packing_dist_d_def}) with projection 2-norm as the distance metric. Such designs correspond to minimizing the bound on the distortion.

Since permutation matrices are special cases of unitary matrices, the limited feedback approach in \cite{Bae_2006_FR-ZFDFE,Jiang_2006_FR-ZFDFE}, in which the precoder is chosen from a codebook of permutation matrices, is a special case of our proposed design. However, the resulting codebooks do not necessarily have the maximum packing distance.
Furthermore, the size of the codebook in the approaches in \cite{Bae_2006_FR-ZFDFE,Jiang_2006_FR-ZFDFE} is fixed for a given $N_t$ and $K$, while the Grassmann packings can be constructed for an arbitrary number of codewords.

\subsection{Codebook Designs for Objectives that are Both Schur-convex and Schur-concave}
For communication objectives  $g (e^\l)$ that are both Schur-convex and Schur-concave functions of $\l$, such as the minimization of product of the MSEs, we observe that the design problem corresponds to maximization of $ {\det (\P^H \H^H \H \P)}/{\sign}$.
Hence, a suitable distortion measure for the codebook is
\begin{align}
\mathcal{{E}}   & =    \mathrm{E}_{\H} \Bigl \{ \det (\bP^{ \text{opt} \: H} \H^H \H  \bP^{\text{opt}}) \notag \\
&\quad\qquad\; -  \det (\bP^{ \text{quant} \: H} \H^H \H  \bP^{\text{quant}})  \Bigr \}/\sign \label{Distortion_def_2} \\
                & =   \mathrm{E}_{\H}\{{\det \Lambdab_{\H 1} }\}/{\sign}
\notag\\
&\quad\qquad\;
          - \mathrm{E}_{\H} \Bigl \{\max_{1 \le j \le |\mathcal{P}|}  \det (\bP^{j \: H} \H^H \H  \bP^{j})  \Bigr \}/{\sign} \nonumber\\
                & \le   \mathrm{E}_{\H}\{{\det \Lambdab_{\H 1} }\}/{\sign}
\notag \\
&\quad
          - \mathrm{E}_{\H}\{{\det \Lambdab_{\H 1} }\}
            \mathrm{E}_{\H} \Bigl \{\max_{1 \le j \le |\mathcal{P}|}  \det (\bP^{j \: H} \U_1 \U_1^H   \bP^{j})  \Bigr \}/ \sign.  \label{Distortion2_bound1}
\end{align}
Here, (\ref{Distortion2_bound1}) follows from the independence of $\U$ and $\boldsymbol{\Lambda}$.
When codebooks are designed from a Grassmann packing using the Fubini-Study distance in (\ref{fubini_dist_def}), the last expectation on the right hand side of (\ref{Distortion2_bound1}) satisfies the following inequality \cite{Love_2005_Grass-LP}:
\begin{equation}
\mathrm{E}_{\H} \Bigl \{\ \max_{1 \le j \le |\mathcal{P}|}  \det (\bP^{j \: H} \U_1 \U_1^H   \bP^{j})  \Bigr \}
\ge
D_{\text{FS}} \cos^2(d_{\text{FS}}/2).
\label{result_from_Love2}
\end{equation}
Hence, we obtain the following upper bound on the distortion:
\begin{equation}
\mathcal{{E}} \le
\mathrm{E}_{\H}\{{\det \Lambdab_{\H 1} }\}
\bigl( 1 - D_{\text{FS}} \cos^2(d_{\text{FS}}/2) \bigr)/ \sign,
\label{Grass_Dist_bound-denisty2}
\end{equation}
which, for a given $|\mathcal{P}|$, is a decreasing function of the packing distance $d_{\text{FS}}$. A similar upper bound was proposed for designing codebooks for MIMO systems with linear receivers \cite{Love_2005_Grass-LP}.
\subsection{Comparison with ZF-Linear Schemes}
In this section, we will show that for a given codebook, the performance of the zero-forcing DFE with limited feedback provides an upper bound on the performance of its linear zero-forcing counterpart for any Schur-convex performance objective $g(e^\l)$. As stated in the following lemma, this is true for any codebook, including those codebooks constructed from non-unitary matrices.
\begin{lem}
Consider a codebook of precoding matrices, $\mathcal{P}$, and a Schur-convex performance  $g(e^\l)$. For any given channel $\H$, let $\l^j_{\text{DFE}}$ denote the vector $\l$ in (\ref{Gras_def_l}) when the precoder $\P^j$ is used, and let the $\l^j_{\text{Lin}}$ denote the corresponding vector for the case of linear equalization. Then
\[
\min_{j = 1, \ldots, |\mathcal{P}|} g(e^{\l^j_{\text{DFE}}})
\le
\min_{j = 1, \ldots, |\mathcal{P}|} g(e^{\l^j_{\text{Lin}}}).
\nonumber
\]
\label{Grass_Lemma3}
\end{lem}

\begin{proof}
Consider a given channel $\H$ and any precoding matrix $\P^j \in \mathcal{P}$. For the linear zero-forcing receiver we have $\C = \I$. It follows from (\ref{DFE_ZF_C}) that the corresponding  matrix $\N^j$  and its Cholesky factor $\L^j$ are diagonal. Hence, $(\Lii^j)^2 = \lambda_i(\N^j)$, or, equivalently,
\[ \l^j_{\text{Lin}} = (\ln \lambda_1(\N^j), \ldots, \ln \lambda_K(\N^j)) . \nonumber \]
On the other hand, for the DFE receiver we have
\[ \l^j_{\text{DFE}} = (\ln (\L_{11}^j)^2, \ldots, \ln (\L_{KK}^j)^2). \nonumber \]
From Lemma~\ref{Grass_Lemma_majorization_ineq}, we have $  \l^j_{\text{DFE}} \prec \l^j_{\text{Lin}}$, hence $g(e^{\l^j_{\text{DFE}}}) \le  g(e^{\l^j_{\text{Lin}}})$ and
\[
\min_{j = 1, \ldots, |\mathcal{P}|} g(e^{\l^j_{\text{DFE}}})
\le
\min_{j = 1, \ldots, |\mathcal{P}|} g(e^{\l^j_{\text{Lin}}}).
\nonumber \]
\end{proof}

\section{Simulation Studies}
In this section, we simulate the performance of the proposed limited feedback MIMO schemes  over a standard i.i.d. Rayleigh block fading channel model.%
\footnote{The coefficients of the channel matrix $\H$ are modelled as independent circularly symmetric complex Gaussian random variables with zero mean and unit variance.}
\rev{For the error rate performance comparisons, we use 16-QAM signaling and} we plot the average  bit error rate (BER) of the $K$ data streams against the signal-to-noise-ratio, which is defined as the ratio of the total average transmitted power $\Ptot$ to the total receiver noise power $\mathrm{E}\{\n^H \n\}$.
We compare the performance of the proposed codebook designs for systems with  zero-forcing DFE  with that of the optimal zero-forcing DFE transceiver for the case of perfect CSI that was presented in Section \ref{Grass_Sec_ZF-Framework}.
For the proposed limited-feedback schemes, the Grassmann codebooks are constructed using the  design approach in \cite{Hochwald_2000_coherentconstellations}; see also \cite{Love_2005_Grass-LP}.
\rev{(Grassmann codebooks could also be constructed using the optimization algorithms in \cite{Edelman_1998_geometry_algo,Manton_2002_Grass})}.
We also provide simulation-based comparisons with the two limited feedback schemes for zero-forcing DFE systems in \cite{Jiang_2006_FR-ZFDFE}.
In addition, we provide performance comparisons  with limited feedback schemes for linear zero-forcing  transceivers that use Grassmann codebooks \cite{Love_2005_Grass-LP}, and with the optimal zero-forcing linear transceiver designs for the case of  perfect CSI  for minimum MSE and minimum bit error rate design criteria \cite{Ding_2003_ZFLin_MinBER}.
%
\subsection{Comparisons with Limited Feedback Zero-forcing DFE Schemes}

\begin{figure}
\centering
\includegraphics[width=3.5in]{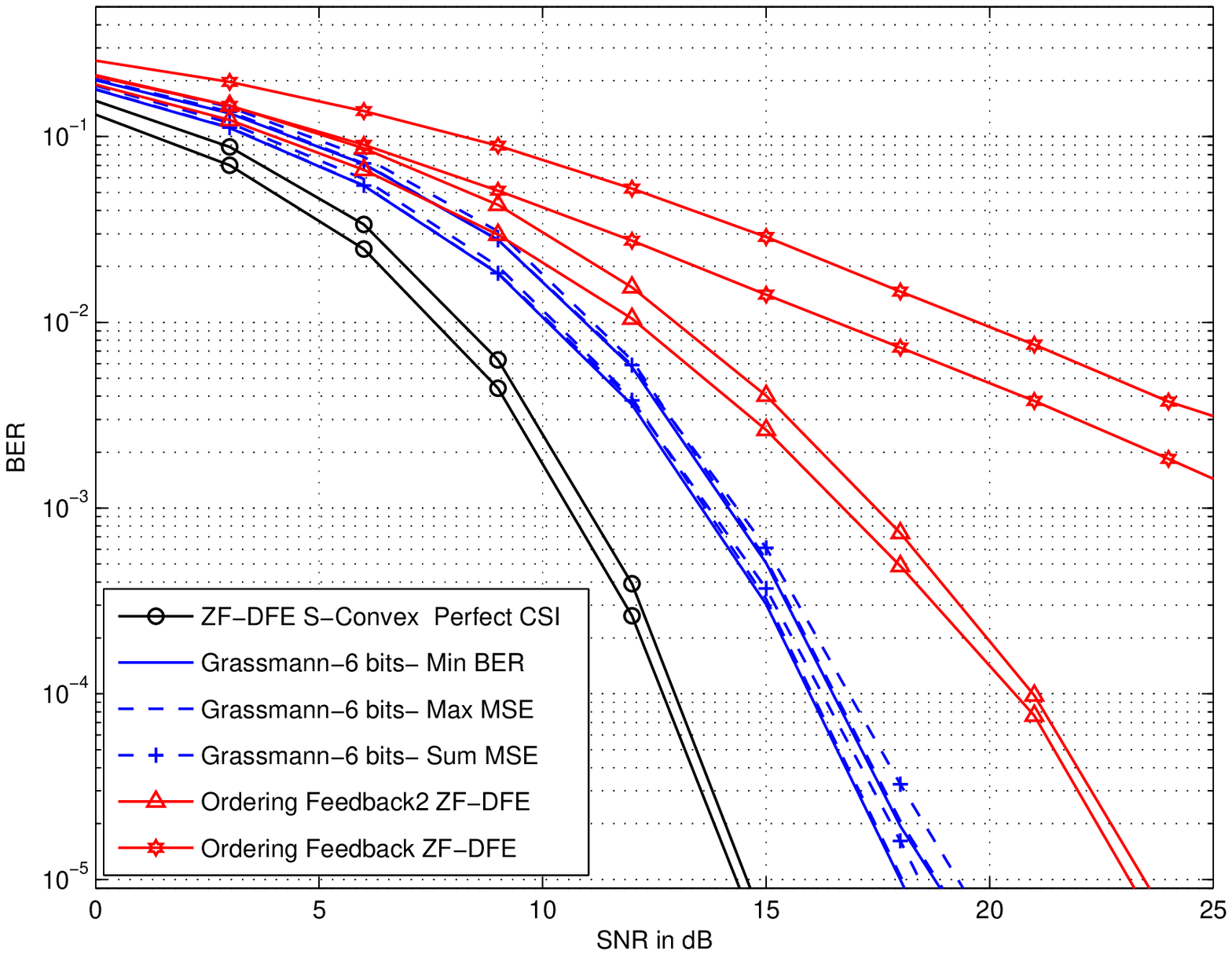}
\caption{ BER performance of various MIMO transmission schemes with   zero-forcing DFE for a system with $N_t = 6, N_r =3$, and $K=3$ simultaneously transmitted 16-QAM data streams.
The schemes considered are: the proposed codebook designs for the objectives of  minimization of the sum of MSEs (Grassmann-6 bits- Sum MSE), minimization of the average BER (Grassmann-6 bits- Min BER); the optimal zero-forcing design  for any  Schur-convex design objective with perfect CSI (ZF DFE - Perfect CSI); and the limited feedback schemes in \cite{Jiang_2006_FR-ZFDFE}, which are based on feeding back the detection ordering (Ordering Feedback - ZF DFE) and (Ordering Feedback2 - ZF DFE).
\rev{The lower curve for each method represents the BER performance obtained under the assumption of correct previous decisions.}}
\label{fig_Grass_16QAM_6_3}
\end{figure}

\begin{figure}
\centering
\includegraphics[width=3.5in]{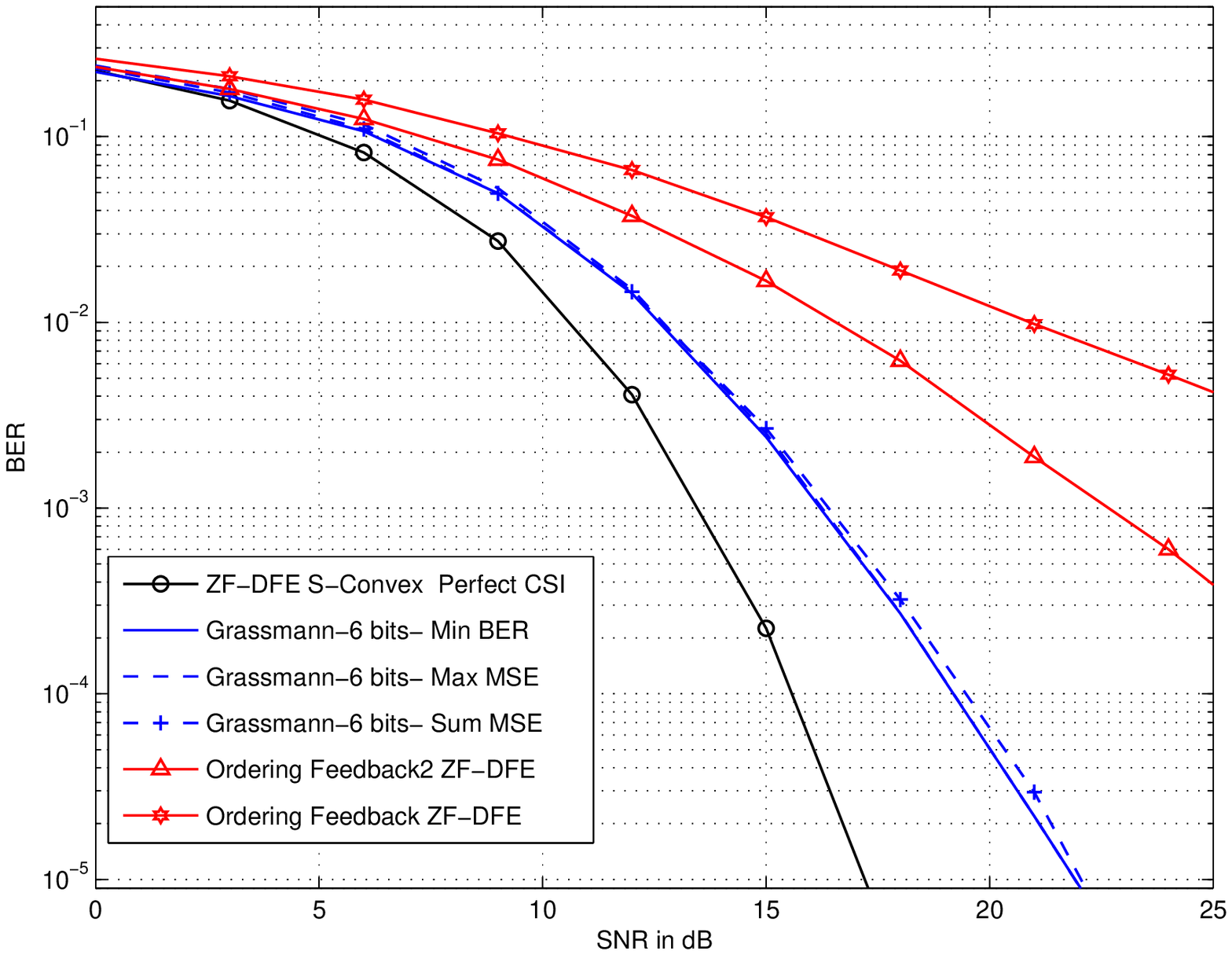}
\caption{ BER performance of various MIMO transmission schemes with   zero-forcing DFE for a system with $N_t = 5, N_r =4$, and $K=4$ simultaneously transmitted 16-QAM data streams.
The schemes considered are: the proposed codebook designs for the objectives of  minimization of the sum of MSEs (Grassmann-6 bits- Sum MSE), minimization of the average BER (Grassmann-6 bits- Min BER); the optimal zero-forcing design  for any  Schur-convex design objective with perfect CSI (ZF DFE - Perfect CSI); and the limited feedback schemes in \cite{Jiang_2006_FR-ZFDFE}, which are based on feeding back the detection ordering (Ordering Feedback - ZF DFE) and (Ordering Feedback2 - ZF DFE).}
\label{fig_Grass_16QAM_5_4}
\end{figure}

\begin{figure}
\centering
\includegraphics[width=3.5in]{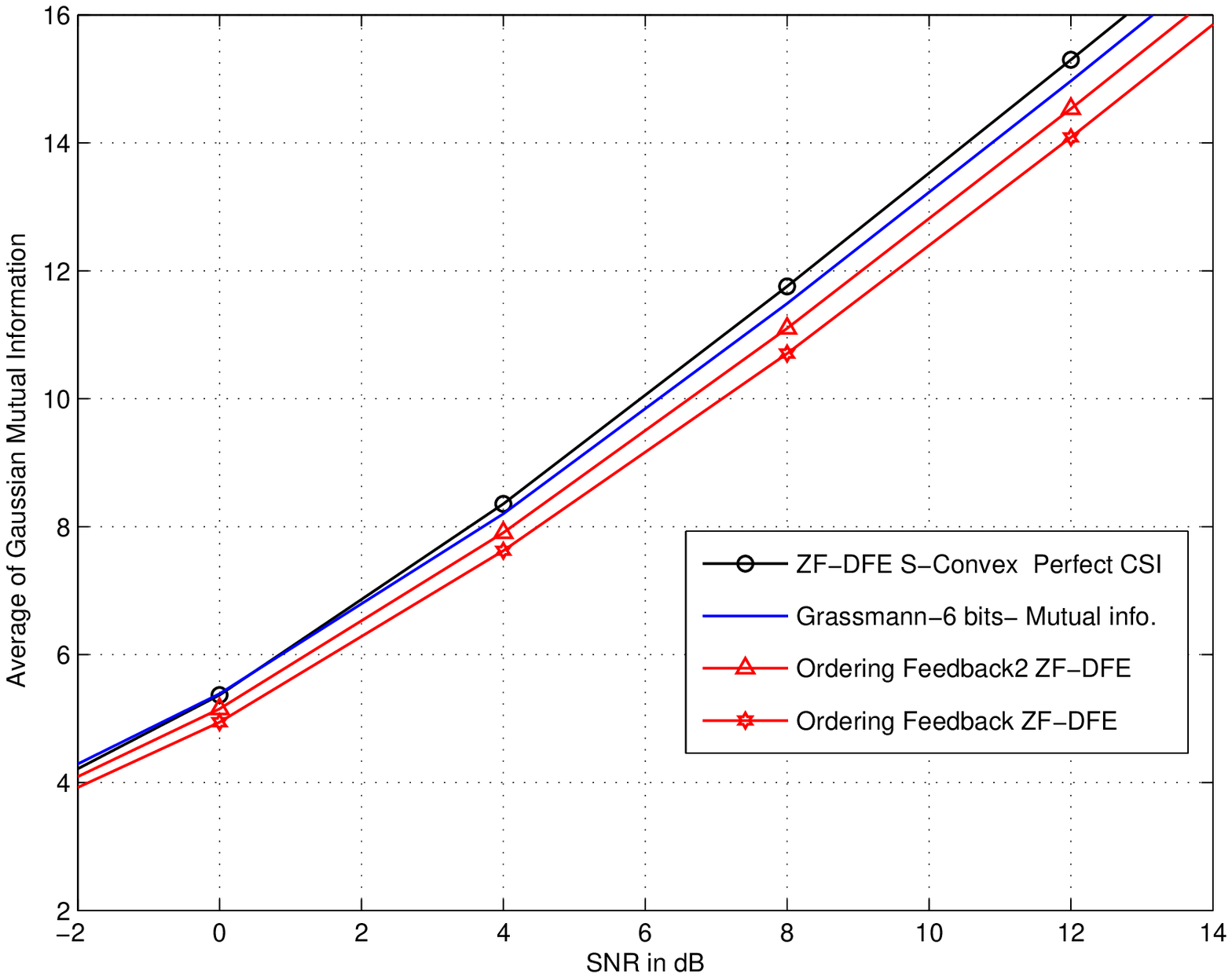}
\caption{ Average of  Gaussian mutual information in (\ref{Grass_GuassianMutualInfo_eg}) for various MIMO transmission schemes with   zero-forcing DFE for a system with $N_t = 5, N_r =4$, and $K=4$.
The schemes considered are: the proposed codebook designs for  Gaussian mutual information objective (Grassmann-6 bits- Mutual info); the optimal zero-forcing design  for any  Schur-convex design objective with perfect CSI (ZF DFE - Perfect CSI); and the limited feedback schemes in \cite{Jiang_2006_FR-ZFDFE}, which are based on feeding back the detection ordering (Ordering Feedback - ZF DFE) and (Ordering Feedback2 - ZF DFE).}
\label{fig_Grass_Cap_5_4}
\end{figure}

In Fig~\ref{fig_Grass_16QAM_6_3}, we consider a MIMO system with $N_t = 6$ transmit antennas and $N_r = 3$ receive antennas that transmits $K = 3$ independent data streams.
We compare the performance of the proposed schemes with Grassmann codebook designs and precoder selection based on the minimization of the sum of the MSEs (Grassmann-6 bits- Sum MSE), minimization of the average BER (Grassmann-6 bits- Min BER), and the minimization of the maximum MSE (Grassmann-6 bits- Max MSE) which is equivalent to  the maximization of minimum SINR.
The codebooks  consist  of 64 unitary matrices, and hence 6 bits of feedback are used per block.
We also make comparisons with the limited feedback schemes in \cite{Jiang_2006_FR-ZFDFE} (Ordering Feedback ZF-DFE  and  Ordering Feedback2 ZF-DFE)  in which the receiver feeds back  the index of the selected permutation of the columns of $\H$  from the set of possible $P^{N_t}_{K} = N_t!/(N_t - K)!$ permutation matrices.
For the system under consideration, the number of possible  permutations  matrices is 120, almost twice the size of the Grassmann codebook.
In the scheme denoted Ordering Feedback ZF-DFE  the permutation matrix is selected based on the norms of the columns of $\H$, while the scheme denoted Ordering Feedback2 ZF-DFE  the permutation is selected based on a greedy ordering of the QR decomposition of the channel matrix $\H$.
In Fig.~\ref{fig_Grass_16QAM_6_3}, we observe the close performance of the proposed codebooks with different Schur-convex selection criteria. This is to be expected,  because in the limit  of infinite feedback~(i.e., perfect CSI), all these objectives result in the same optimal precoder design.
We also observe that the Grassmann codebooks provide significantly better performance than the schemes that are based on precoding with permutation matrices, even though they employ  fewer feedback bits. This is because codebooks constructed from permutation matrices are special cases of those constructed from unitary matrices, and they do  not necessarily minimize the distortion measures.
\rev{
Note that for all error performance figures in this paper, the simulation results of all ZF-DFE methods include the effect of error propagation.
For reference, in Fig~\ref{fig_Grass_16QAM_6_3} we also provide the  performance under the assumption of correct previous decisions; i.e., no error propagation.
We observe that at high SNRs, the practical performance of the optimal zero-forcing DFE transceiver for the case of perfect CSI and the proposed designs based on Grassmann codebooks are close to their corresponding performance in absence of error propagation.
This also holds for the permutation feedback scheme (Ordering Feedback2 ZF-DFE).
}

Analogous performance advantages to those in Fig~\ref{fig_Grass_16QAM_6_3} are observed in Fig~\ref{fig_Grass_16QAM_5_4}, which shows the performance for a MIMO system with $N_t = 5$ transmit antennas and $N_r = 4$ receive antennas that transmits $K = 4$ data streams.
The size of each permutation-based codebook  is 120 matrices, while the size of each Grassmann codebook  is 64 matrices.


\rev{In Fig~\ref{fig_Grass_Cap_5_4} we compare several different methods in terms of the Gaussian mutual information that they achieve. We consider a system with  $N_t = 5$, $N_r = 4$, and $K=4$, and we plot the average, over 1000 channel realizations, of the Gaussian mutual information achieved by the ZF-DFE transceiver with the quantized precoder; i.e., the average of the values of (\ref{Grass_GuassianMutualInfo_eg}) achieved by the quantized precoder.
For the proposed scheme we consider a Grassmann codebook design  and precoder selection based on the maximization  of the Gaussian mutual information   (Grassmann-6 bits- Mutual info.), and a codebook that consists  of 64 unitary matrices.
We make comparisons with the limited feedback schemes in \cite{Jiang_2006_FR-ZFDFE} (Ordering Feedback ZF-DFE  and  Ordering Feedback2 ZF-DFE), whose permutation-based codebooks contain 120 matrices.
We observe that the proposed Grassmann codebook with precoder selection based on the maximization of the Gaussian mutual information provides the closest performance to the optimal ZF-DFE design for the case of perfect CSI, which was presented in Section \ref{Grass_Sec_ZF-Framework}.
}
\subsection{Comparisons with Limited Feedback Linear Zero-forcing  Schemes}
\begin{figure}
\centering
\includegraphics[width=3.5in]{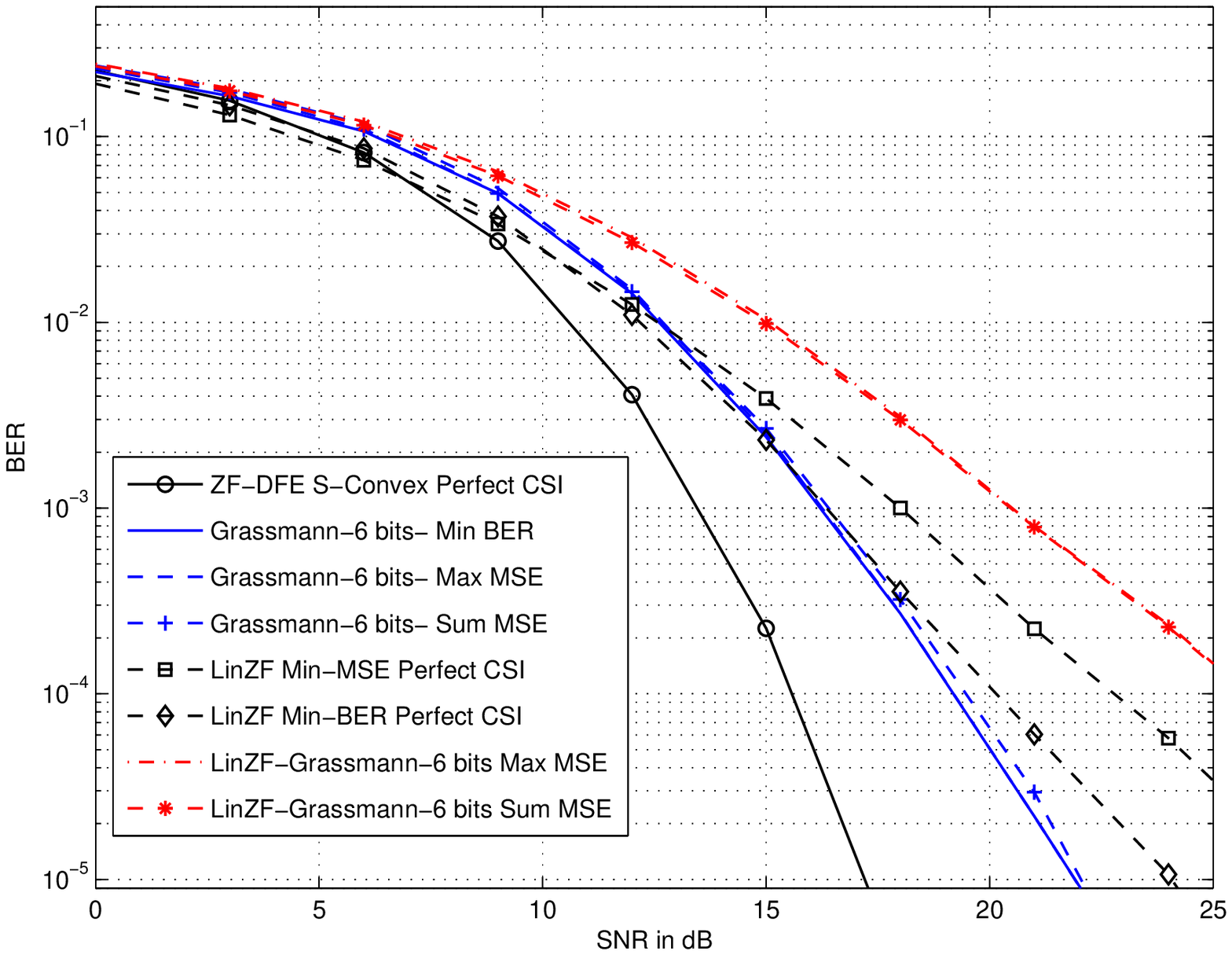}
\caption{ BER performance of various MIMO transmission schemes with   zero-forcing linear and DFE systems with $N_t = 5, N_r =4$, and $K=4$ simultaneously transmitted 16-QAM data streams.
The schemes considered are: the proposed codebook designs for the objectives of  minimization of the sum of MSEs (Grassmann-6 bits- Sum MSE), minimization of the average BER (Grassmann-6 bits- Min BER); the optimal zero-forcing design  for any  Schur-convex design objective with perfect CSI (ZF DFE - Perfect CSI);  the optimal linear zero-forcing design  for minimum MSE (LinZF Min-MSE Perfect CSI) and minimum average BER (LinZF Min-BER Perfect CSI) \cite{Ding_2003_ZFLin_MinBER}; and the linear zero-forcing limited feedback schemes in \cite{Love_2005_Grass-LP} for minimum total MSE (LinZF-Grassmann-6 bits Sum MSE) and minimum maximum MSE (LinZF-Grassmann-6 bits Max MSE).}
\label{fig_Grass_16QAM_5_4_Lin}
\end{figure}

\begin{figure}
\centering
\includegraphics[width=3.5in]{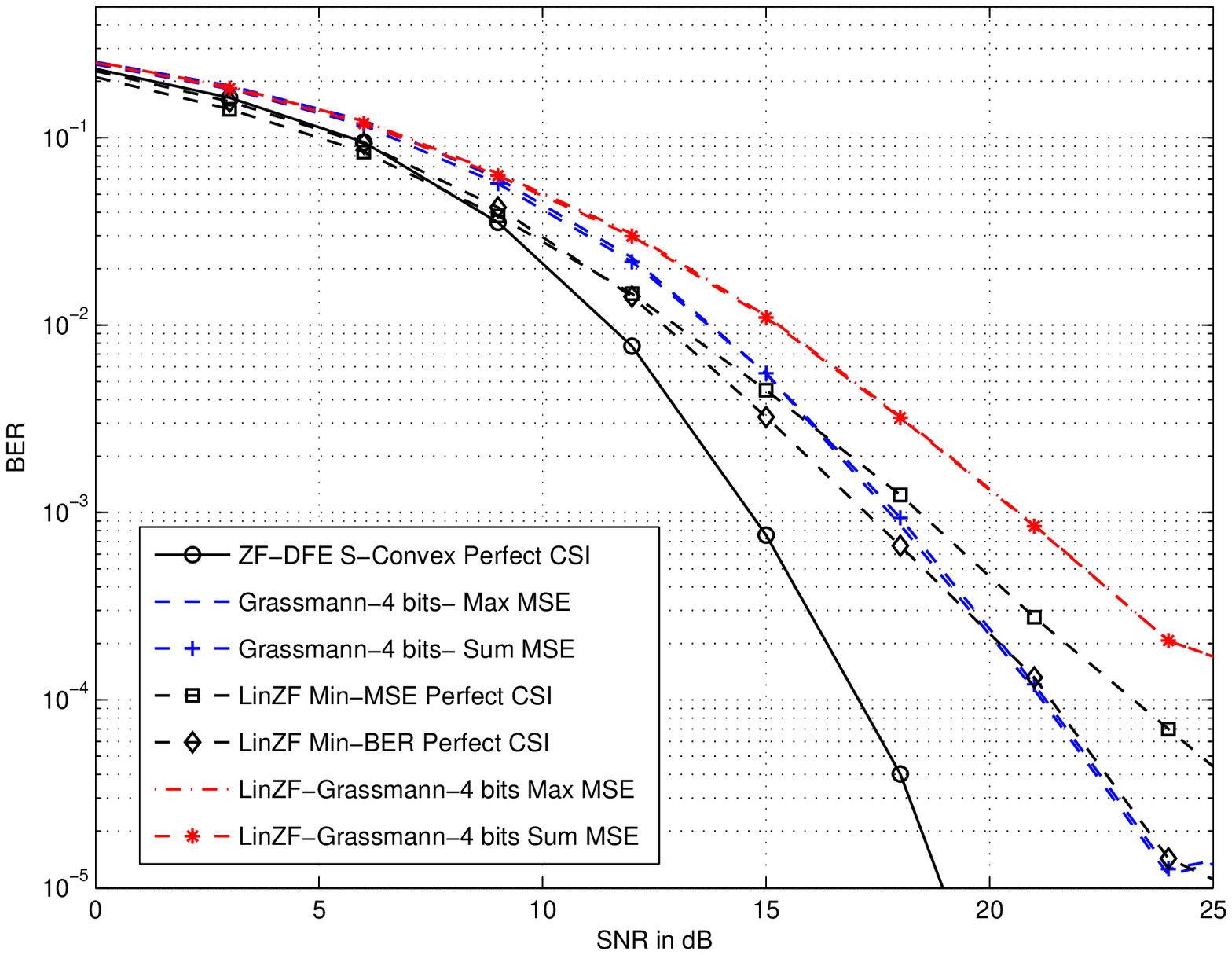}
\caption{ BER performance of various MIMO transmission schemes with   zero-forcing linear and DFE systems with $N_t = 4, N_r =3$, and $K=3$ simultaneously transmitted 16-QAM data streams.
The schemes considered are: the proposed codebook designs for the objectives of  minimization of the sum of MSEs (Grassmann-6 bits- Sum MSE), minimization of the average BER (Grassmann-6 bits- Min BER); the optimal zero-forcing design  for any  Schur-convex design objective with perfect CSI (ZF DFE - Perfect CSI);  the optimal linear zero-forcing design  for minimum MSE (LinZF Min-MSE Perfect CSI) and minimum average BER (LinZF Min-BER Perfect CSI) \cite{Ding_2003_ZFLin_MinBER}; and the linear zero-forcing limited feedback schemes in \cite{Love_2005_Grass-LP} for minimum total MSE (LinZF-Grassmann-6 bits Sum MSE) and minimum maximum MSE (LinZF-Grassmann-6 bits Max MSE).}
\label{fig_Grass_16QAM_4_3_Lin}
\end{figure}

In Fig~\ref{fig_Grass_16QAM_5_4_Lin}, we consider a MIMO system with $N_t = 5$ transmit antennas and $N_r = 4$ receive antennas that transmits $K = 4$ independent data streams.
We compare the performance of the proposed ZF-DFE schemes that use Grassmann codebooks with that of the corresponding linear zero-forcing schemes that use Grassmann codebooks with the same feedback rate \cite{Love_2005_Grass-LP}.
We consider  linear limited feedback schemes with different precoder selection criteria, namely minimization of the total MSE  (LinZF-Grassmann-6 bits Sum MSE), and maximization of the minimum eigen value of the overall channel $\H \P$ (LinZF-Grassmann-6 bits Max MSE), which corresponds to minimization of the maximum MSE \cite{Love_2005_Grass-LP}.
We also provide performance comparisons with the zero-forcing DFE transceiver design  for perfect CSI that simultaneously optimizes any Schur-Convex design criteria, and  with the corresponding optimal  zero-forcing linear transceiver designs  for perfect CSI that minimize the total MSE or the average BER. Unlike the DFE case, these two  design criteria  result in different precoder designs \cite{Ding_2003_ZFLin_MinBER}.
In Fig.~\ref{fig_Grass_16QAM_5_4_Lin}, we observe that the proposed zero-forcing  DFE systems with limited feedback perform better than the corresponding linear schemes; as is to be expected, c.f.   Lemma~\ref{Grass_Lemma3}.
Similar performance advantages are observed in Fig~\ref{fig_Grass_16QAM_4_3_Lin} for a MIMO system with $N_t = 4$ transmit antennas and $N_r = 3$ receive antennas that transmits $K = 3$ independent data streams.


\section{Conclusion}
We have considered the design of  multiple-input multiple-output communication systems with zero-forcing decision feedback equalization (DFE) when only limited rate feedback from the receiver to the transmitter is  available.
We  considered schemes in which the receiver uses its CSI to select the best available precoder from a codebook of precoders and then feeds back the index of this precoder to the transmitter using a small number of  bits.
To facilitate the development of the limited feedback scheme, we developed a unified design framework for the joint design of  the precoder and DFE receiver when perfect channel state information is available.
We then  characterized the statistical distribution of the optimal precoder in a standard Rayleigh fading environment, and
showed that codebooks constructed from Grassmann packings minimize an upper bound on an average distortion measure.
%
%
Our simulation studies showed that the proposed limited feedback scheme can provide significantly better performance with a lower feedback rate than the existing schemes in which the detection order is fed back to the transmitter.

\appendix
\section{Proof of Theorem 1}
\label{Grass_Proof_Theorem1}
\subsection{ Optimal Precoder for Schur-convex Functions}
If $g(e^\l)$ is a Schur convex function of $\l$, then from Lemma~\ref{Grass_Lemma_majorization_ineq} we have that
\begin{equation}
g(e^{\lb}) \le g(e^\l),
\end{equation}
and the optimal value is obtained when all $\l_i$ are equal to
\begin{equation}
\l_i = \frac{1}{K} \ln \det (\N).
\end{equation}
Hence, all MSEs are equal to $\E_{ii} = \Lii^2 = \sqrt[K]{\det(\N)}$.
Since the objective is an increasing function of the individual MSEs, the design goal
reduces to minimizing $\det \N$ subject to the power constraint on the precoder and to the constraint that
diagonal elements of the Cholesky  factor of $\N$ are all equal.
We will start by characterizing the family of precoders that minimize $\det(\N)$ subject
to the power constraint, then we will show that there is a member of this family that
yields a Cholesky factor of $\N$ with equal diagonal elements.
Minimizing $\det(\N)$ is equivalent to maximizing \rev{$\det (\P^H \H^H \H \P)$}, and the family of
optimal precoders is given by \cite{Botros_2007_VTC}:
\begin{equation}
\P = \sqrt{\frac{\Ptot}{K}} \U_1  \V,
\label{ZF_precoder_family}
\end{equation}
where $\U_1 \in \mathbb{C}^{N_t \times K}$ contains the eigen vectors of  $\H^H \H$
corresponding to the $K$ largest eigen values, and $\V \in \mathbb{C}^{K \times K}$ is a
unitary matrix degree of freedom.
To complete the design of $\P$, we need to  select $\V$ such that the
Cholesky~decomposition of $\N = \L \L^H$ yields an $\L$ factor with equal diagonal
elements.
Using (\ref{ZF_precoder_family}) we have that
\begin{eqnarray}
\N   & =  &  \frac{K \sign}{\Ptot}
         \left( \V^H  \Lambdab^{-1/2}_{\H 1}      \right)
         \left(       \Lambdab^{-1/2}_{\H 1}   \V \right) \nonumber \\
     & =  &  \L \L^H = \R^H \R = (\Q \R)^H(\Q \R),
\end{eqnarray}
where $\Lambdab_{\H 1}$ is the diagonal matrix containing the largest $K$ eigen values
of $\H^H \H$, and $\Q$ is a matrix with orthonormal columns.
Therefore, finding $\V$ is equivalent to finding a $\V$ such that QR~decomposition of
$\bigl(       \Lambdab^{-1/2}_{\H 1}   \V \bigr)$ has an R-factor with equal diagonal.
This problem was solved in \cite{Zhang_2005_QRS,Jiang_2005_GMD_math}, and $\V$  can be obtained by applying the
algorithms therein  to the matrix  $\Lambdab^{-1/2}_{\H 1}$.

\subsection{Optimal Precoder for Schur-concave Functions}
If $g(e^\l)$ is a Schur-concave function of $\l$, then from Lemma~\ref{Grass_Lemma_majorization_ineq}  we have that
$g(e^\l)$  is minimized when   $\Lii^2  =  \lambda_i(\N)$,
and that this  equality holds when $\L$ is normal matrix.
Since $\L$ is a lower triangular matrix,  in order for it to be normal it must be a diagonal
matrix \cite{MAtrix_Analysis}.
The optimal $\C$ in that case is $\I$, and hence $\B = \0 $.
That is, in the case of Schur-concave functions of $\l$, the optimal  ZF-DFE  design
results in zero-forcing linear equalization.


\begin{IEEEbiography}{Michael Botros Shenouda}
received the B.Sc.\ (Hons.\ 1) degree in 2001 and the M.Sc.\ degree in 2003, both in electrical
engineering and both from Cairo University, Egypt.
He is currently working toward the Ph.D. degree at the Department of Electrical and Computer Engineering, McMaster University, Canada.
His main areas of interest include wireless and MIMO communication, convex and robust optimization, and signal processing algorithms.
He is also interested in majorization theory, and its use in the development of design frameworks for non-linear MIMO transceivers.
Mr.\ Botros Shenouda was awarded an IEEE Student Paper Award at ICASSP 2006,
and was a finalist in the IEEE Student Paper Award competition at ICASSP 2007.
\end{IEEEbiography}

\begin{IEEEbiography}{Tim Davidson} (M'96) received the B.Eng.\ (Hons.~I) degree in Electronic Engineering from the University of Western Australia (UWA), Perth, in 1991 and the D.Phil. degree in Engineering Science from the University of Oxford, U.K., in 1995.

He is currently an Associate Professor in the Department of Electrical and Computer Engineering at McMaster University, Hamilton, Ontario, Canada, where he holds the (Tier II) Canada Research Chair in Communication Systems, and is currently serving as Acting Director of the School of Computational Engineering and Science. His research interests lie in the general areas of communications, signal processing and control. He has held research positions at the Communications Research Laboratory at McMaster University, the Adaptive Signal Processing Laboratory at UWA, and the Australian Telecommunications Research Institute at Curtin University of Technology, Perth, Western Australia.

Dr.\ Davidson was awarded the 1991 J. A. Wood Memorial Prize (for
``the most outstanding [UWA] graduand'' in the pure and applied
sciences) and the 1991 Rhodes Scholarship for Western Australia. He
is currently serving as an Associate Editor of the IEEE Transactions
on Signal Processing and as an Editor of
the IEEE Transactions on Wireless Communications.
He has also served as an Associate Editor of
the IEEE Transactions on Circuits and
Systems II, and as a Guest Co-editor of issues of
the IEEE Journal on Selected Areas in Communications and the IEEE
Journal on Selected Topics in Signal Processing.
\end{IEEEbiography}
\end{document}